\documentstyle[11pt,a4,cite,epsf]{article}
\newcommand{\lesssim}{ {\
\lower-1.2pt\vbox{\hbox{\rlap{$<$}\lower5pt\vbox{\hbox{$\sim$}}}}\ }  }
\newcommand{\gtrsim}{ {\
\lower-1.2pt\vbox{\hbox{\rlap{$>$}\lower5pt\vbox{\hbox{$\sim$}}}}\ }  }

\newcommand{\be}{\begin{equation}}
\newcommand{\ee}{\end{equation}}
\newcommand{\bea}{\begin{eqnarray}}
\newcommand{\eea}{\end{eqnarray}}
\newcommand{\noi}{\noindent}
\newcommand{\nn}{\nonumber}

\newcommand{\cF}{{\cal F}}

\newcommand{\cM}{{\cal M}}

\newcommand{\cO}{{\cal O}}

\newcommand{\cW}{{\cal W}}

\newcommand{\Imm}{\mbox{\rm Im}}
\newcommand{\QCD}{\mbox{\rm {\tiny QCD}}}

\newcommand{\MeV}{\mbox{\rm MeV}}
\newcommand{\GeV}{\mbox{\rm GeV}}
\newcommand{\fm}{\mbox{\rm fm}}

\newcommand{\with}{\mbox{\rm with}}
\newcommand{\annd}{\mbox{\rm and}}

\begin{document}

\begin{titlepage}

\begin{flushright} CPT-97/P.3519 \\UB-ECM-PF 97/15. \\ \today
\end{flushright}
\vspace*{2cm}
\begin{center} {\Large \bf How Small Can the Light Quark Masses Be?}\\[1.5cm]
{\large {\bf L.~Lellouch}$^{a}$, {\bf E.~de Rafael}$^{a}$, and  {\bf
J.~Taron}$^{b}$}\\[1cm] $^a$ Centre  de Physique Th\'eorique\\ CNRS-Luminy,
Case 907\\ F-13288 Marseille Cedex 9, France\\[0.3cm] and\\[0.3cm]
${}^b$ Departament d'Estructura i Constituents de la Mat\`{e}ria\\
Facultat de F\'{\i}sica, Universitat de Barcelona, and IFAE\\
Diagonal 647, E-08028 Barcelona, Spain\\
\end{center}

\vspace*{1.0cm}
\begin{abstract}

We derive lower bounds for the combination of light quark masses $m_s +m_u$
and $m_d +m_u$. The derivation is based on first principles: the
analyticity properties of two--point functions of local current operators
and the
positivity of the corresponding hadronic spectral functions. The bounds follow
from the restriction of the sum over all possible hadronic states which can
contribute to the hadronic spectral functions to those with the
lowest invariant mass. These hadronic contributions are well known
phenomenologically, either from experiment or from chiral perturbation theory
calculations. The results we find cast serious doubts on the
reliability of some of the lattice QCD determinations which have been recently
reported.

\end{abstract}

\vfill


\end{titlepage}

{\large{\bf 1 Introduction.}}
\vspace*{3 mm}

\noi
There are recent determinations of the light quark masses from lattice
QCD~\cite{BG96,Mactal96} which find substantially lower values than those
previously  obtained using a variety of QCD sum rules, (see e.g.
refs.~\cite{BPR95,JM95,CDPS95} and earlier references therein.) The new
lattice determinations are also in disagreement with earlier lattice
results~\cite{Alletal94}. More recently, an independent lattice QCD
determination~\cite{Eietal97} using dynamical Wilson fermions finds results
which, for $m_{u}+m_{d}$ are in agreement within errors with those of
refs.~\cite{BG96,Mactal96}; while for
$m_{s}$ they agree rather well, again within errors, with the QCD sum rule
determinations. The present situation concerning the values of
the light quark masses is therefore rather confusing.

The purpose of this note is to show that there exist rigorous
lower bounds on how small the light quark masses $m_{s}+m_{u}$
and $m_{d}+m_{u}$ can be. The derivation of the bounds is based on first
principles: the analyticity properties of two--point functions of local
current operators and the positivity of the corresponding hadronic spectral
functions. Analyticity relates integrals of the hadronic spectral functions to
the behaviour of QCD two--point functions in the deep euclidean region via
dispersion relations. The bounds follow from the restriction of the sum over
all possible hadronic states which can contribute to the spectral function to
the state(s) with the lowest invariant mass. It turns out that, for the
two--point functions which we shall consider, these hadronic contributions are
well known phenomenologically; either from experiment or from chiral
perturbation theory ($\chi$PT) calculations. On the QCD side of the dispersion
relation, the two--point functions in question where the quark masses appear
as an overall factor, are known in the deep euclidean region from
perturbative QCD (pQCD) at the four loop level. The leading
non--perturbative power corrections which appear in the operator product
expansion in the physical vacuum~\cite{SVZ79} are also known. As we shall show,
the bounds we find cast serious doubts on the reliability of the recent lattice
determinations reported in refs.~\cite{BG96,Mactal96,Eietal97}.

We shall be concerned with two types of local operators, one is
the divergence of the strangeness changing axial current
\be
\partial_{\mu}A^{\mu}(x)=(m_{s}+m_{u}):\!\!{\bar s}(x)i\gamma_{5}u(x)\!\!:\,,
\ee
with its corresponding two--point function
\be\label{eq:2pf5}
\Psi_{5} (q^2)=i\int d^{4}x e^{iq\cdot x}\langle 0\vert
T\left(\partial_{\mu}A^{\mu}(x)\partial_{\nu}A^{\nu}
(0)^{\dagger}\right)\vert 0\rangle\,;
\ee
the other one is the scalar operator $S(x)$, defined as the isosinglet
component
of the mass term in the QCD Lagrangian
\be S(x)=\hat{m}[:\!\!{\bar u}(x)u(x)\!\!:+:\!\!{\bar
d}(x)d(x)\!\!:], \qquad
\mbox{\rm where}\qquad \hat{m}\equiv\frac{1}{2}(m_u + m_d)\,,
\ee
and its corresponding two--point function
\be\label{eq:2sf5}
\Psi (q^2)=i\int d^{4}x e^{iq\cdot x}\langle 0\vert
T\left(S(x)S
(0)^{\dagger}\right)\vert 0\rangle\,.
\ee
We  shall refer to these two--point functions respectively  as
the ``pseudoscalar channel'' and the ``scalar channel'', and
discuss them separately.
\goodbreak

{\large{\bf 2 The Pseudoscalar Channel.}}
\vspace*{3 mm}

\noi
The function  $\Psi_{5}(q^2)$ in (\ref{eq:2pf5}) obeys a dispersion
relation and in QCD it requires two subtractions. It is then appropriate to
consider its second derivative $(Q^2=-q^2)$:
\be\label{eq:disp}
\Psi_{5}''(Q^2)=\int_{0}^{\infty}
dt\frac{2}{(t+Q^2)^3}\frac{1}{\pi}\Imm\Psi_{5}(t)\,,
\ee
with
\be
\frac{1}{\pi}\Imm\Psi_{5}(t)=\sum_{\Gamma}\langle0\vert
\partial_{\mu}A^{\mu}(0)\vert\Gamma
\rangle\langle\Gamma\vert\partial_{\mu}A^{\mu}(0)^{\dagger}\vert 0\rangle
(2\pi)^{3}
\delta^{(4)}(q-\sum p_{\Gamma})\,.
\ee
The sum over $\Gamma$ extends to all possible  on--shell hadronic states
with the quantum numbers of the operator $\partial_{\mu}A^{\mu}$ and it defines
the corresponding hadronic spectral function.  The behaviour of the
$\Psi_{5}''(Q^2)$ function at large euclidean values of the external momentum
squared $Q^2=-q^2 \gg \Lambda_{QCD}^2$ is known from pQCD
calculations~\cite{BNRY81,S89,GKLS90,GKLS91,C96}:
\be
\Psi_{5}''(Q^2)=\frac{N_c}{8\pi^2}\frac{\left[m_{s}(Q^2)+m_{u}(Q^2)\right]
^2}{Q^2}\left[1+
\frac{11}{3}\frac{\alpha_{s}(Q^2)}{\pi} +\cdots\right]\,,
\ee
where the dots represent higher order terms which have been calculated
up to
${\cal O}(\alpha_s^3)$~\cite{C96}, as well as  non--perturbative
power corrections of
$\cO\left(\frac{1}{Q^4}\right)$~\cite{BNRY81,PR82},
and strange quark mass corrections of
$\cO\left(\frac{m_{s}^2}{Q^2}\right)$ and
$\cO\left(\frac{m_{s}^4}{Q^4}\right)$ including ${\cal O}(\alpha_s)$
terms~\cite{Ge90,SCh88,ChGS85,JM95}.

It is convenient to consider moments of the
two--point function
$\Psi_{5}(q^2)$ as follows:
\be\label{eq:moment}
\cM_{5}^{(N)}(Q^2)=\frac{2(-1)^N}{(N+2)!}(t_0 +Q^2)^{N}
\frac{\partial^N}{(\partial Q^2)^N}\Psi_{5}''(Q^2)\,,\qquad N=0,1,2,\dots\,,
\ee
and explicitly separate the contribution from the
$K$--pole contribution to the spectral function in the dispersive
integral in eq.~(\ref{eq:disp}). We then have
\be\label{eq:sigmaN}
\cM_{5}^{(N)}(Q^2)-4f_{K}^{2}M_{K}^{4}
\frac{(t_{0}+Q^2)^N}{(M_{K}^{2}+Q^2)^{N+3}}=\Sigma_{N}(Q^2)\,,
\ee
where $\Sigma_{N}(Q^2)$ denotes the hadronic continuum integral [$t_{0}\equiv
(M_{K}+2m_{\pi})^2$]
\be\label{eq:contint}
\Sigma_{N}(Q^2)=\int_{t_{0}}^{\infty}dt\frac{2}{(t+Q^2)^3}
\left(\frac{t_0 +Q^2}{t+Q^2}\right)^{N}\frac{1}{\pi}\Imm\Psi_{5}(t)\,.
\ee
We are now confronted with a typical moment problem~\footnote{See
e.g. ref.~\cite{AhK62}.}. The positivity of the continuum spectral function
$\frac{1}{\pi}\Imm\Psi_{5}(t)$ constrains the moments
$\Sigma_{N}(Q^2)$ and hence the l.h.s. of (\ref{eq:sigmaN}) where the light
quark masses appear. The most general constraints among the first three moments
for
$N=0,1,2$ are:
\be
\Sigma_{0}(Q^2)\ge 0,\qquad \Sigma_{1}(Q^2)\ge 0,\qquad
\Sigma_{2}(Q^2)\ge 0\,;
\ee
\be\label{eq:diff}
\Sigma_{0}(Q^2)-\Sigma_{1}(Q^2)\ge 0,\qquad
\Sigma_{1}(Q^2)-\Sigma_{2}(Q^2)\ge 0\,;
\ee
\be\label{eq:quad}
\Sigma_{0}(Q^2)\Sigma_{2}(Q^2)-\left(\Sigma_{1}(Q^2)\right)^2\ge 0\,.
\ee
The inequalities in eq.~(\ref{eq:diff}) are in fact trivial unless $2Q^2<
t_{0}$, which defines a region in $Q^2$  where unfortunately pQCD is not
applicable. The other inequalities lead however to interesting bounds which we
next discuss.~\footnote{Lower bounds on light quark masses based on
inequalities of this type were first discussed in ref.~\cite{BNRY81}. Our
discussion here is an update of this work.}

\vspace*{3 mm}
{\it 2a. The Bounds from $\Sigma_{N}(Q^2)\ge 0$.}

\noi
The inequality $\Sigma_{0}(Q^2)\ge 0$ results in a first bound on the
running masses:
\be\label{eq:1stbound}
\left[m_{s}(Q^2)+m_{u}(Q^2)\right]^2 \ge \frac{16\pi^2}{N_c}
\frac{2f_{K}^2 M_{K}^4}{Q^4}\frac{1}{\left(1+\frac{M_{K}^2}{Q^2}\right)^3}
\frac{1}{\left[1+
\frac{11}{3}\frac{\alpha_{s}(Q^2)}{\pi} +\cdots\right]}\,.
\ee
The bound can of course be converted into a bound of the same combination of
quark masses at any fixed reference scale using the renormalization group
running of quark masses. The favourite choice of the lattice community is
$4\:\GeV^2$. We shall adopt it here as well so as to
make easier the comparison of our bounds with their results.
Notice that the bound in eq.~(\ref{eq:1stbound}) is non--trivial in the
large--$N_c$ limit ($f_{K}^2\sim\cO(N_c)$) and in the chiral limit
($m_{s}\sim M_{K}^2$). The bound is of course a function of the choice of the
euclidean $Q$--value at which the r.h.s. in eq.~(\ref{eq:1stbound}) is
evaluated. For the bound to be meaningful, the choice of $Q[\GeV]$ has to be
made sufficiently large so that a pQCD evaluation of the QCD factor
\be\label{eq:qcdf0}
\cF_{0}^{\QCD}(Q^2)=1+\frac{11}{3}\frac{\alpha_{s}(Q^2)}{\pi}+\cdots\,,
\ee
associated to the two--point function $\cM_{5}^{(0)}(Q^2)$ in
eq.~(\ref{eq:moment}) is applicable. We consider that, $Q\gtrsim 1.4\:\GeV$
is a safe choice. The evaluation of $\cF_{0}^{\QCD}$ at $Q=1.4\:\GeV$ with or
without inclusion of the ${\cal O}(\alpha_s^3)$ terms differs by less than
10\%. The lower bound which follows from eq.~(\ref{eq:1stbound}) for
$m_u + m_s$ at a renormalization scale
$\mu^2=4\:\GeV^2$
results in the solid curves shown in Fig.~1 below. These are
the lower bounds obtained by letting $\Lambda^{(3)}_{\overline{MS}}$
vary~\cite{PDB} between 290~MeV (the upper curve) and 380~Mev (the lower curve)
and using $f_{K}=113\:\MeV$, $M_{K}=493.67\:\MeV$. We take the number of active
flavours
$n_f=3$ in all the numerical analyses of this work. Notice that the hadronic
continuum integral in (\ref{eq:contint}) is always larger than the contribution
from three light flavours; and indeed, we have checked that using
$n_{f}=4$ in the QCD expressions leads to even higher bounds. Values of the
quark
masses below the solid curves in Fig.~1 are forbidden by the bounds. For
comparison, the horizontal lines in the figure correspond to the central values
obtained by the authors of ref.~\cite{BG96}: their ``quenched'' result is the
horizontal upper line; their ``unquenched'' result the horizontal lower line.

It should
be emphasized that in Fig.~1 we are comparing what is meant to be a
``calculation'' of the quark masses --the horizontal lines which are the lattice
results of ref.~\cite{BG96}-- with a bound  which in fact can only be saturated
in the limit where
$\Sigma_{0}(Q^2)= 0$. Physically, this limit corresponds to the extreme case
where the hadronic spectral function from the continuum of states is
totally neglected! What the plots show is that, even in that extreme limit,
and  for values of $Q$ in the range
$1.4\:\GeV\lesssim Q\lesssim 1.8\:\GeV$ the new lattice results are already in
difficulty with the bounds.

\vskip 2pc
\centerline{\epsfbox{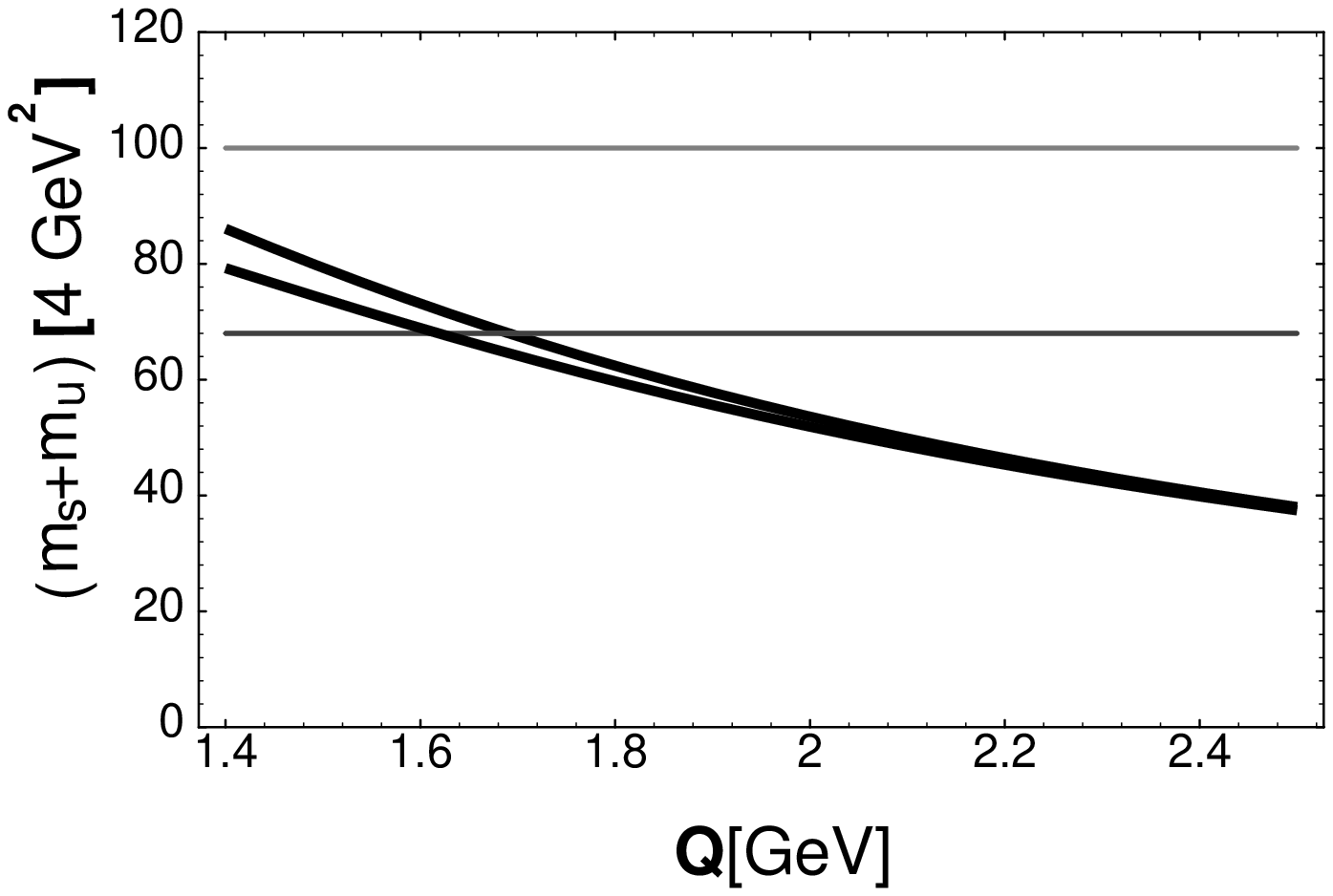}}
\vspace*{3 mm}
\centerline{{\bf Fig.~1} {\it Lower bound in {\rm MeV} for
$[m_{s}+m_{u}](4\:\GeV^2)$ versus $Q(\GeV)$ from eq.~(\ref{eq:1stbound}).}}
\vskip 2pc

The inequalities $\Sigma_{1}(Q^2)\ge 0$ and $\Sigma_{2}(Q^2)\ge 0$  result in
the improved bounds
\be\label{eq:2ndbound}
\left[m_{s}(Q^2)+m_{u}(Q^2)\right]^2 \ge \frac{16\pi^2}{N_c}
\frac{2f_{K}^2 M_{K}^4}{Q^4}\frac{3}{\left(1+\frac{M_{K}^2}{Q^2}\right)^4}
\frac{1}{\left[1+
\frac{17}{3}\frac{\alpha_{s}(Q^2)}{\pi} +\cdots\right]}\,;
\ee
and

\be\label{eq:3rdbound}
\left[m_{s}(Q^2)+m_{u}(Q^2)\right]^2 \ge \frac{16\pi^2}{N_c}
\frac{2f_{K}^2 M_{K}^4}{Q^4}\frac{6}{\left(1+\frac{M_{K}^2}{Q^2}\right)^5}
\frac{1}{\left[1+
\frac{20}{3}\frac{\alpha_{s}(Q^2)}{\pi} +\cdots\right]}\,.
\ee
The improvement is due to the fact that higher and higher moments increase the
weight of the contribution from the low energy part of the hadronic
spectrum, in our case the $K$--pole; e.g., at $Q=2\:\GeV$
and for $\Lambda^{(3)}_{\overline{MS}}=330\:\MeV$ eq.(\ref{eq:2ndbound}) implies
$(m_{s}+m_{u})[4\:\GeV^2]\ge 77\:\MeV$ and eq.(\ref{eq:3rdbound})
$(m_{s}+m_{u})[4\:\GeV^2]\ge 106\:\MeV$. Notice however that as the bounds are
improved by going to higher moments, the corresponding QCD functions
\be\label{eq:qcdf12}
\cF_{1}^{\QCD}(Q^2) =  1+
\frac{17}{3}\frac{\alpha_{s}(Q^2)}{\pi}+ \cdots\,,\qquad
\cF_{2}^{\QCD}(Q^2) =  1+
\frac{20}{3}\frac{\alpha_{s}(Q^2)}{\pi}+ \cdots\,,
\ee
have larger coefficients in their pQCD expansion in powers of $\alpha_{s}$.
Therefore, for the improved bounds to be meaningful, the appropriate choices of
$Q$ have to be made at sufficiently larger values so that a pQCD evaluation of
the $\cF_{N}^{\QCD}$--functions is applicable.
The same phenomenon
occurs when one considers the quadratic bound in eq.(\ref{eq:quad}) which
we next discuss. However, as we shall see, the net improvement at the
appropriate higher $Q$--values is still quite considerable.

\goodbreak

{\it 2a. The Bounds from the Quadratic Inequality.}

\noi
Inserting eq.~(\ref{eq:sigmaN}) in the quadratic inequality in
eq.~(\ref{eq:quad}) and solving for the quark masses results in the bound
\bea\label{eq:qbound}
\lefteqn{\left[m_{s}(Q^2)+m_{u}(Q^2)\right]^2 \ge
\frac{16\pi^2}{N_c}
\frac{2f_{K}^2
M_{K}^4}{Q^4}\frac{9}{\left(1+\frac{M_{K}^2}{Q^2}\right)^5}\times}
\nn \\ & & \times\frac{2\cF_{0}^{\QCD}(Q^2) -\frac{4}{3}
\left(1+\frac{M_{K}^2}{Q^2}\right)\cF_{1}^{\QCD}(Q^2)
+\frac{1}{3}\left(1+\frac{M_{K}^2}{Q^2}\right)^2\cF_{2}^{\QCD}(Q^2)}
{3\cF_{0}^{\QCD}(Q^2)\cF_{2}^{\QCD}(Q^2)
-2\left(\cF_{1}^{\QCD}(Q^2)\right)^2}\,,
\eea
where $\cF_{0}^{\QCD}(Q^2)$, $\cF_{1}^{\QCD}(Q^2)$, and $\cF_{2}^{\QCD}(Q^2)$ are
the QCD factors (\ref{eq:qcdf0}) and (\ref{eq:qcdf12}) associated to the
two--point functions
$\cM_{5}^{(N)}$ in eq.(\ref{eq:moment}) with $N=0,1,2$:
\be\label{eq:qcdc}
3\cF_{0}^{\QCD}\cF_{2}^{\QCD} -2\left(\cF_{1}^{\QCD}\right)^2=
1+\frac{25}{3}\frac{\alpha_{s}}{\pi}+\cdots \,.
\ee
The resulting lower bounds for the quark masses are the solid lines shown in
Fig.~2 below. Like in Fig.~1, these are the lower bounds obtained by letting
$\Lambda^{(3)}_{\overline{MS}}$ vary~\cite{PDB} between 290~MeV (the upper
curve)
and 380~Mev (the lower curve). Values of the quark masses below these solid
curves are forbidden by the bounds. The horizontal lines in this figure are also
the same lattice results as in Fig.~1.

\vskip 2pc
\centerline{\epsfbox{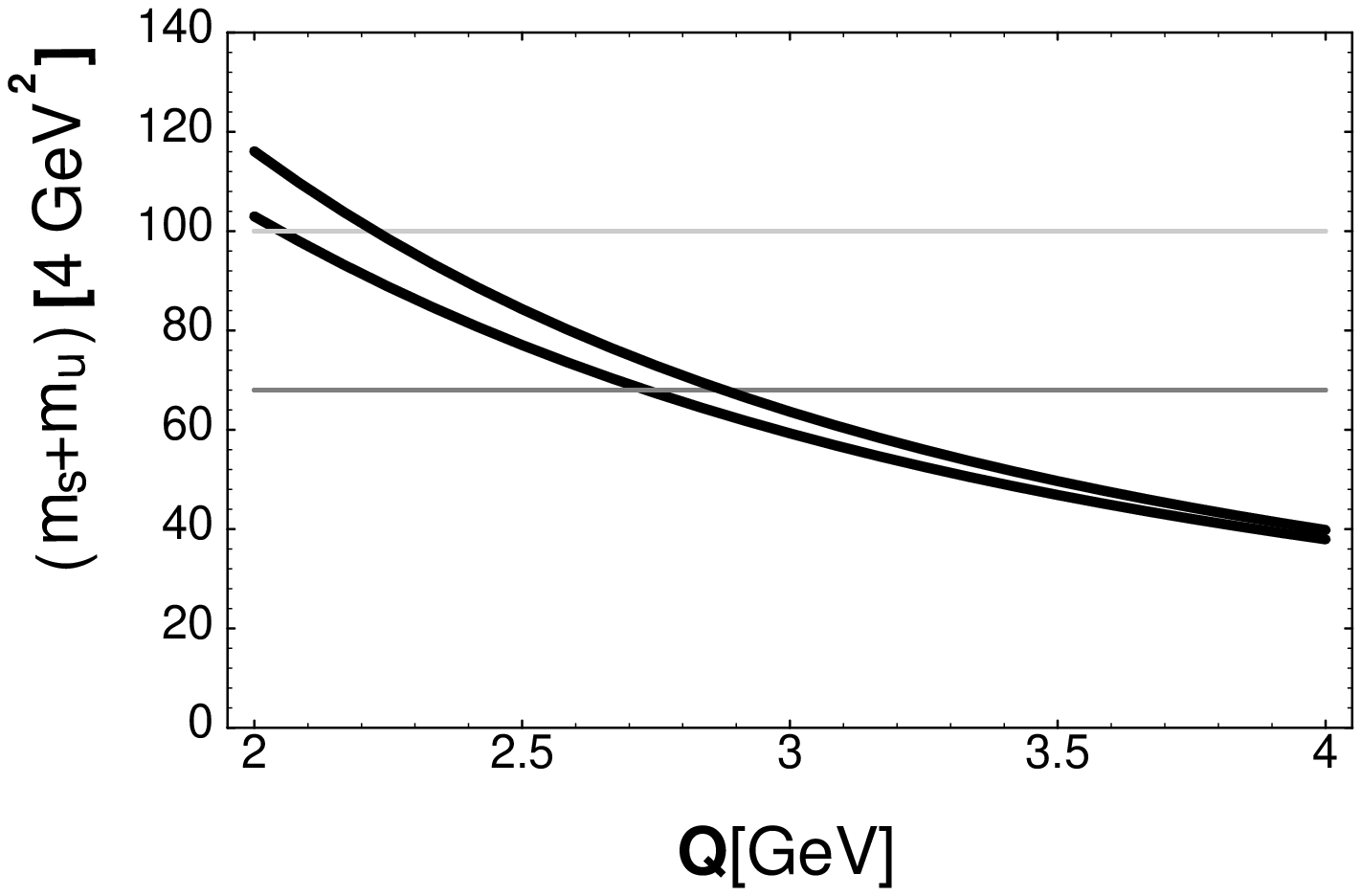}}
\vspace*{3 mm}
\centerline{{\bf Fig.~2} {\it Lower bound in {\rm MeV} for
$[m_{s}+m_{u}](4\:\GeV^2)$ from the quadratic inequality in
eq.(\ref{eq:qbound}).}}
\vskip 2pc

\noi
The quadratic bound in eq.(\ref{eq:qbound}) is saturated when
$\Sigma_{0}(Q^2)\Sigma_{2}(Q^2)-\left(\Sigma_{1}(Q^2)\right)^2=0$, which
happens for a $\delta$--like spectral function representation of the
hadronic continuum of states at an arbitrary position and with an
arbitrary weight. This is certainly less restrictive than the extreme limit with
the full hadronic continuum neglected, and it is therefore not surprising that
the quadratic bound happens to be better than the ones from
$\Sigma_{N}(Q^2)$ for
$N=0,1,$ and $2$. Notice that the quadratic bound in Fig.~2 is plotted at higher
$Q$--values than the bound in Fig.~1. The evaluation of the QCD factor in
(\ref{eq:qcdc}) with or without inclusion of the
$\cO(\alpha_{s}^{3})$ terms differs by less than 10\% for
$Q^2\ge 4\:\GeV^2$  and we consider
therefore that for the evaluation of the  quadratic bound $Q\gtrsim 2\:\GeV$ is
already a safe choice. We find that even the quenched lattice results
of refs.~\cite{BG96,Mactal96} are in difficulty with these bounds.

Similar bounds can be obtained for $m_{u}+m_{d}$ if instead of the two--point
function in (\ref{eq:2pf5}) one considers the two--point function associated
to the divergence of the axial current
\be
\partial_{\mu}A^{\mu}(x)=(m_{d}+m_{u}):\!\!{\bar d}(x)i\gamma_{5}u(x)\!\!:\,.
\ee
The method to derive the bounds is exactly the same as the one discussed
above and therefore we only show, in Fig.~3 below, the results for the
corresponding lower bounds  which we obtain from the quadratic inequality.

\vskip 1pc
\centerline{\epsfbox{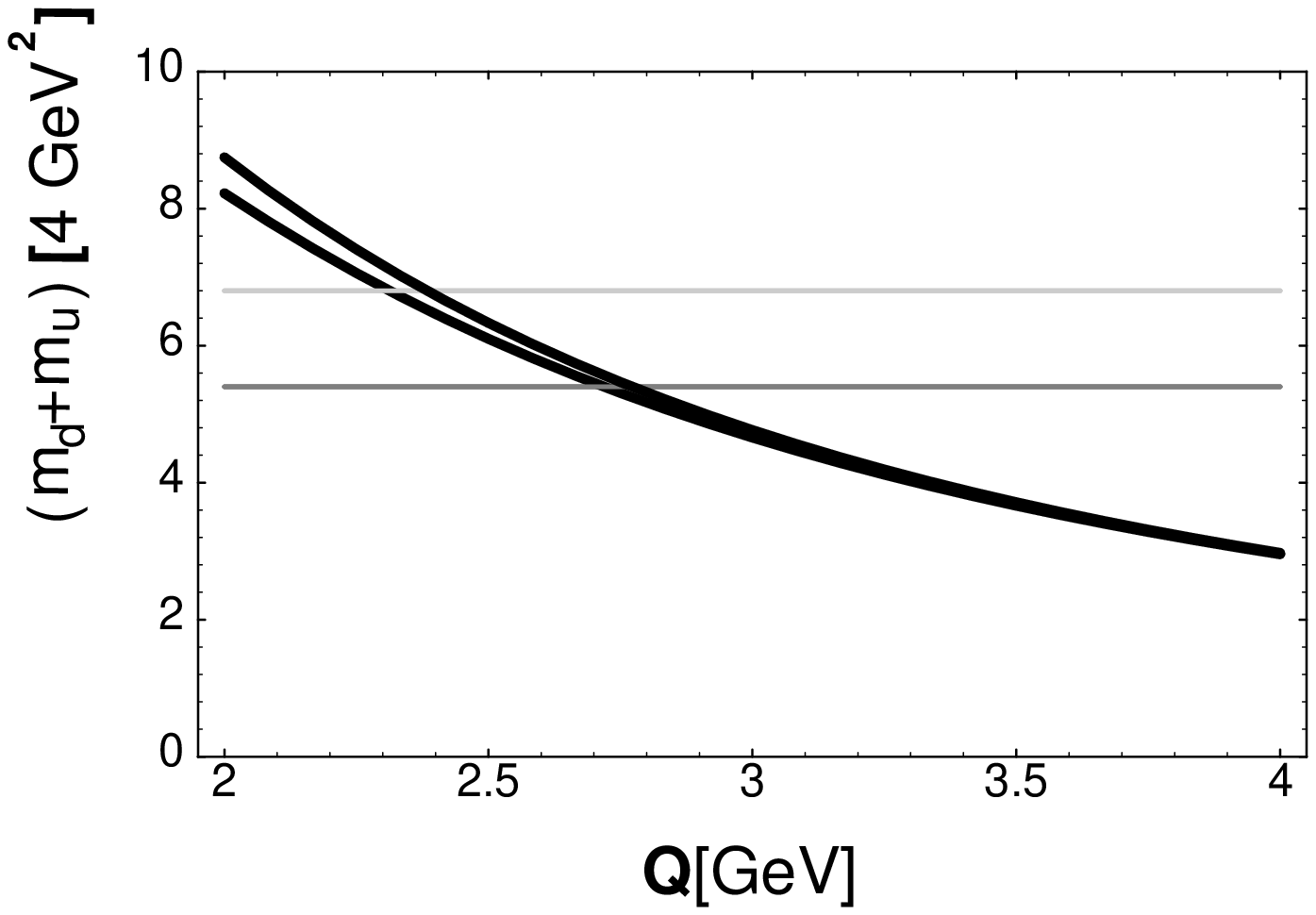}}
\vspace*{3 mm}
\centerline{{\bf Fig.~3} {\it Lower bound in {\rm MeV} for
$[m_{d}+m_{u}](4\:\GeV^2)$ from the quadratic inequality in
eq.(\ref{eq:qbound}).}}
\vskip 2pc

\noi
We find again that the lattice QCD results of
refs.~\cite{BG96,Mactal96,Eietal97} for $m_{u}+m_{d}$ are in serious
difficulties with these bounds. Notice that in this case the mass corrections
in the QCD two--point function are negligible at the plotted $Q$--values.

\goodbreak


{\large{\bf 3 The Scalar Channel.}}
\vspace*{3 mm}

\noi
The function  $\Psi(q^2)$ in eq.~(\ref{eq:2sf5}) obeys a dispersion relation
which in QCD requires two subtractions. We shall thus work with its second
derivative
\be\label{eq:disps}
\Psi''(Q^2)=\int_{0}^{\infty}
dt\frac{2}{(t+Q^2)^3}\frac{1}{\pi}\Imm\Psi(t)\,.
\ee
The QCD behaviour of the
$\Psi''(Q^2)$ function at large euclidean values of the external momentum
squared $Q^2=-q^2 \gg \Lambda_{QCD}^2$ is entirely similar to
the one of $\Psi_{5}''(Q^2)$ which we have already discussed, but for very
small $m_u$ and $m_d$ quark mass correction terms  which we now neglect  i.e.,
\be
\Psi''(Q^2)=2\times\frac{N_c}{8\pi^2}\frac{\hat{m}^{2}(Q^2)}{Q^2}\left[1+
\frac{11}{3}\frac{\alpha_{s}(Q^2)}{\pi} +\cdots\right]\,.
\ee
The difference between the ``scalar'' and ``pseudoscalar'' channels is
that the lowest state which contributes to the
hadronic spectral function in the ``scalar'' channel is not a pole but the
$\pi-\pi$ continuum with
$J=0$ and $I=0$. This contribution provides a lower bound to the full
spectral function:
\be\label{eq:specin}
\frac{1}{\pi}\Imm\Psi(t)\ge\frac{1}{16\pi^2}\sqrt{1-\frac{4m_{\pi}^2}
{t}}\,3\,\vert F(t)\vert^{2}\theta(t-4m_{\pi}^2)\,,
\ee where $F(t)$ denotes the scalar--isoscalar pion form factor
\be
\langle \pi^{a}(p)\pi^{b}(p')\vert S(0)\vert 0\rangle =\delta_{ab}
F(t)\,,\qquad \with   \qquad t=(p+p')^2\,.
\ee
In particular, the value of this form factor at the origin is the so
called {\it pion sigma term}. This form factor has been well studied in
$\chi$PT at the one and two loop level (see e.g. refs\cite{DGL90,GM91} and
references therein), and the predicted low energy shape agrees well with the
available experimental information.
With
\be\label{eq:formf}
F(t)=F(0)\left[1+\frac{1}{6}\langle r^{2}\rangle_{s}^{\pi}t+
\cO(t^2)\right]\,,
\ee
it is found~\cite{GM91}
that~\footnote{The pion mass used in ref.~\cite{GM91} is the $\pi^{+}$--mass.
Since the electromagnetic interactions are neglected, we find it more
appropriate to use the $\pi^{0}$--mass instead; which in any case results in
a lower contribution to the average quark mass. In the generalized
version of $\chi$PT (see e.g. ref.~\cite{KMS95} and references therein,) the
value of
$F(0)$ could be sensibly larger, which would result in even larger bounds. A
value for the curvature of the scalar form factor is also quoted in
ref.~\cite{GM91}; it involves however some extra phenomenological assumptions
which go beyond
$\chi$PT and this is why we have not included this extra information here.}
\be
F(0)=m_{\pi}^{2}\times (0.99\pm 0.02) + \cO(m_{\pi}^{6})
 \qquad \annd \qquad \langle r^{2}\rangle_{s}^{\pi}= (0.59\pm 0.02)~\fm^{2}
\ee

Using standard methods~\footnote{see ref.~\cite{deRT94} and references therein,
and also refs.\cite{AMS75,AEMS75,Mi73}} we construct bounds on
$\hat{m}$ when something is known about the scalar--isoscalar pion form factor
$F(t)$. For this, it is convenient to map the complex $q^2$--plane to
the complex unit disc:
\be i\frac{1+z}{1-z}=\sqrt{y-1}\qquad {\mbox{\rm with}} \qquad
y=\frac{t}{4m_{\pi}^2}\,.
\ee
The dispersion relation in (\ref{eq:disps}) and the spectral function
inequality in (\ref{eq:specin}) can then be combined as follows:
\be\label{eq:masteriny}
\cW(Q^2)\ge\frac{1}{\pi}\int_{1}^{\infty} dy \left(y+\frac{Q^2}
{4m_{\pi}^2}\right)^{-3}y^{-1/2}(y-1)^{1/2}\vert F(4m_{\pi}^2 y)\vert^2
\equiv
\int_{-\pi}^{+\pi}\frac{d\theta}{2\pi}\vert\phi(e^{i\theta})F(t(e^{i\theta})
)\vert^2\,,
\ee
where $\phi(z)$ is an analytic function on the unit disc which contains the
phase--space factors as well as the dispersive weight factor and the jacobian of
the conformal mapping; and
\be
\cW(Q^2)\equiv \frac{16\pi}{3}\times (4m_{\pi}^{2})^2
2\frac{N_c}{16\pi^2}\frac{\hat{m}^{2}(Q^2)}{Q^2}\left[1+
\frac{11}{3}\frac{\alpha_{s}(Q^2)}{\pi}+\cdots\right]\,.
\ee

With $F(0)$ and $F'(0)$ as input, the analiticity of
$\phi(z)F(z)$ on the unit disc enables us to write the
constraint~\cite{deRT94}
\be
\cW(Q^2)\ge\frac{2}{\left(1+\sqrt{1+\frac{Q^2}{4m_{\pi}^2}}\right)^{6}}
F^{2}(0)+
\left[F'(0)\frac{dt}{dz(0)}\phi[z(0)]
+F(0)\frac{d\phi(z)}{dz(0)}\right]^2 \,,
\ee
which results in a {\it scalar channel bound} for $\hat{m}^2$:
\be\label{eq:bounds}
\hat{m}^2(Q^2)\ge
\frac{\pi}{N_c}\frac{m_{\pi}^2}{4}
\left(\frac{4m_{\pi}^2}{Q^2}\right)^2\frac{3\times (0.99\pm0.02)^{2}}
{\left(\sqrt{1+\frac{4m_{\pi}^2}{Q^2}}+\frac{2m_{\pi}}{Q} \right)^6}
\frac{1+\left\{-\frac{1}{2} +3z[-Q^2]-
\frac{8m_{\pi}^2}{3}\langle r^{2}\rangle_{s}^{\pi}
\right\}^2}{\left[1+\frac{11}{3}\frac{\alpha_{s}(Q^2)}{\pi}+\cdots
\right]}\,.
\ee
Notice that this bound is somewhat similar to the {\it first bound}
obtained from
the divergence of the axial current in eq.~(\ref{eq:1stbound}). There are
however
some interesting differences which we wish to point out. From the point of view
of the large--$N_c$ expansion, the r.h.s. of eq.~(\ref{eq:bounds}) is
$1/N_c$--suppressed. Also,  from the point of view of the chiral expansion the
r.h.s. of eq.~(\ref{eq:bounds}) has an extra factor $m_{\pi}^2$ as compared to
the r.h.s. of eq.~(\ref{eq:1stbound}). One might expect therefore that the lower
bound in (\ref{eq:bounds}) should be much worse than the one in
(\ref{eq:1stbound}). Yet the bound for $\hat{m}$ in eq.(\ref{eq:bounds}) is
surprisingly competitive. The reason for this is twofold. On the one hand, the
absence of a prominent narrow resonance in the
$J=0$, $I=0$ channel, is compensated by a rather large contribution from
the chiral loops, a phenomenological feature which appears to be common in all
the explored physical processes where the  $J=0$,
$I=0$ channel contributes. On the other hand the fact that, numerically,
$m_{\pi}^2$ and $f_{\pi}^2$ are not so different.
The lower bounds for the sum of running masses
$(m_{u}+m_{d})$ at the renormalization scale $[\mu^2=4\:\GeV^2]$ which
follow from
this {\it scalar channel bound} are the dashed curves shown in Fig~4,
corresponding to the choice
$\Lambda_{\overline{MS}}^{(3)}=290\:\MeV$ (the upper dashed curve) and
$\Lambda_{\overline{MS}}^{(3)}=380\:\MeV$ (the lower dashed curve).
The plots show that for values of $Q$ in the range $1.4\:\GeV\lesssim
Q\lesssim 1.6\:\GeV$ the lattice QCD results
in~\cite{BG96,Mactal96,Eietal97} are
dangerously close to lower bounds obtained with very little phenomenological
input. Notice also that the relevant two--point function in this channel is
practically the same as
$\cM_{5}^{(0)}(Q^2)$ in eq.~(\ref{eq:moment}) for which
$Q\gtrsim 1.4\:\GeV$ is already a safe choice.

One can still improve these bounds by taking into account the extra information
provided by the fact that the phase of the scalar form factor $F(t)$ in the
elastic region $4m_{\pi}^2\le t \le 16m_{\pi}^2$ is precisely the $I=0$, $J=0$
$\pi -\pi$ scattering phase--shift and there is information on this phase--shift
both from $\chi$PT and from experiment.

\centerline{\epsfbox{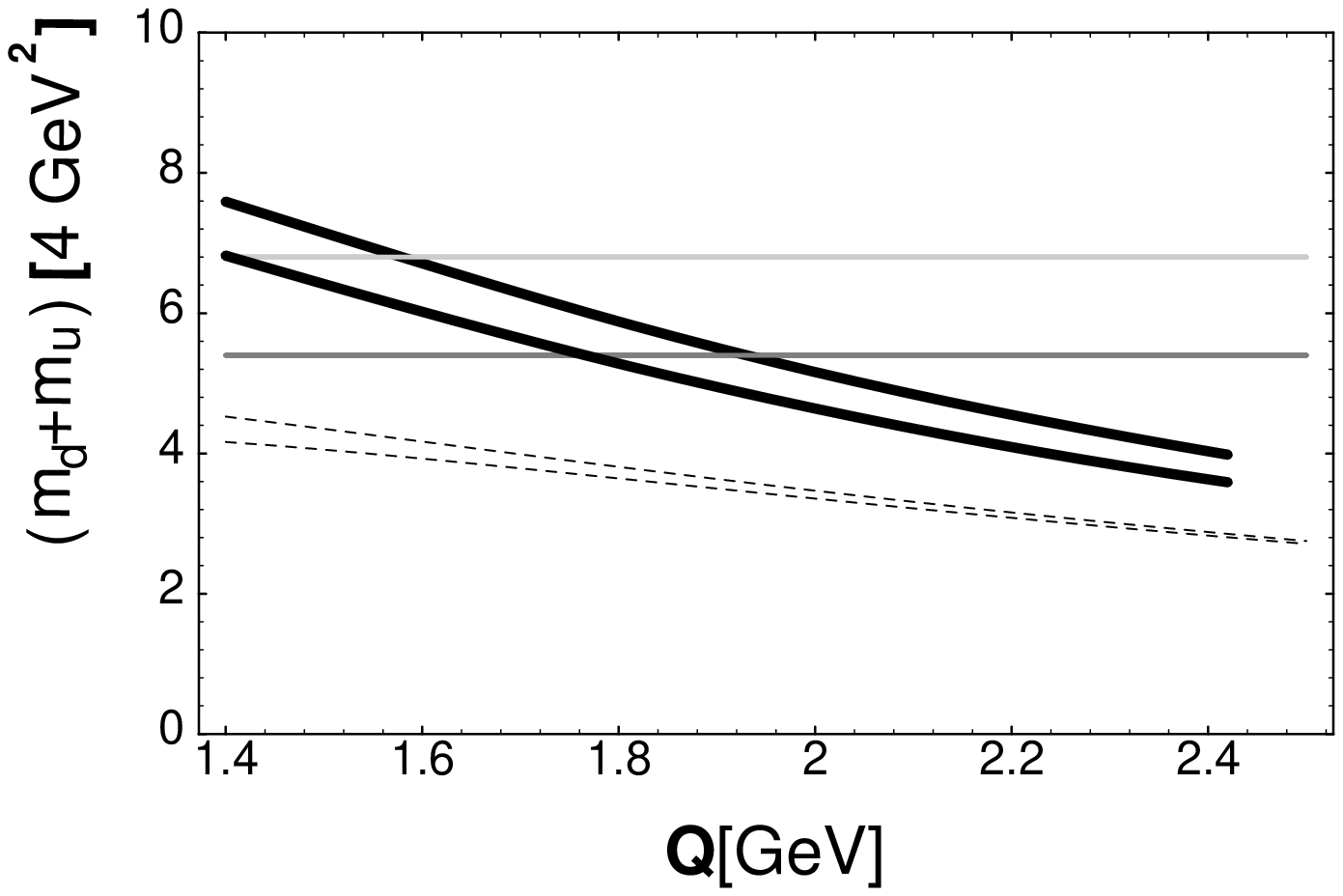}}
\vspace*{3mm}
\centerline{{\bf Fig.~4} {\it Lower bounds in {\rm MeV} for
$[m_{d}+m_{u}](4\:\GeV^2)$ from the Scalar Channel.}}
\vskip 2pc

\noi
The technology to
incorporate this information is discussed in refs.\cite{AMS75,AEMS75,Mi73}. We
have restricted the phase--shift input to a conservative region: $4m_{\pi}^2
\le t \le 500\:\MeV^2$, where a resummation of the chiral logarithms calculated
at the two--loop level in $\chi$PT~\cite{DGL90,GM91} is certainly expected to be
reliable and is in good agreement with experiment. This results in improved
bounds
which correspond to the solid curves shown in Fig.~4. The two curves reflect the
variation within the quoted errors of the input parameters. Again, we find that
the lattice determinations of refs.~\cite{BG96,Mactal96,Eietal97} do not satisfy
lower bounds which only incorporate  very little hadronic input.

\vspace*{7mm}
{\large{\bf 4 Conclusions.}}
\vspace*{3 mm}

We conclude that the lattice results of refs.~\cite{BG96,Mactal96,Eietal97} are
in serious difficulties with rigorous lower bounds which can be derived from
general properties of QCD and with a minimum of well established
phenomenological
input; and this in two different hadronic channels. The lower bounds we have
derived are perfectly compatible with the sum rules results of
refs.~\cite{BPR95,JM95,CDPS95} and earlier references therein.~\footnote{There
exists a recent preprint~\cite{BGM97} with various criticisms on the
determination of quark masses with QCD sum rules. A reply to these criticisms
will be made elsewhere~\cite{PdeR}.}  They are also compatible with a recent
semiempirical determination of $m_{s}$ made by the ALEPH
collaboration~\cite{ALEPH} based on the rate of the hadronic $\tau$--decays into
strange particles observed at LEP.

\goodbreak
\vspace*{7mm}
{\large{\bf Acknowledgments:}}
\vspace*{3 mm}

\noi
We are very grateful to Marc Knecht and Ximo Prades for several helpful
discussions on topics related to light quark masses. J.T. acknowledges
financial suppot from CICYT under contract number AEN95-0590 and from CIRIT
under contract number GRQ93-1047.



\end{document}